\documentclass[preprint]{aastex7}
\usepackage{amsmath}
\usepackage{subcaption}
\usepackage{graphicx}
\usepackage{float}
\usepackage{placeins}

\definecolor{forestgreen}{rgb}{0.10, 0.80, 0.10}

\newcommand{\ganjz}[1]{#1}

\begin{document}

\title{A Monte Carlo simulation on the scattering coefficients of solar radio wave propagation}

\shorttitle{Monte Carlo simulation on scattering of radio waves }
\shortauthors{Gan \& Wang}

\author{Jiazhen Gan}
\affiliation{CAS Key Laboratory of Geospace Environment, School of Earth and Space science, \\
University of Science and Technology of China, Hefei 230026, China}
\affiliation{CAS Center for Excellence in Comparative Planetology, \\ University of Science and Technology of China, Hefei 230026, China}
\email{ganjz07@mail.ustc.edu.cn}  

\author[0000-0001-6252-5580]{Chuanbing Wang} 
\affiliation{CAS Key Laboratory of Geospace Environment, School of Earth and Space science, \\
University of Science and Technology of China, Hefei 230026, China}
\affiliation{CAS Center for Excellence in Comparative Planetology, \\ University of Science and Technology of China, Hefei 230026, China}
\email[show]{cbwang@ustc.edu.cn}


\begin{abstract}

Radio waves undergo scattering by small-scale density fluctuations during propagation through the solar-terrestrial environment, substantially affecting the observed characteristics of solar radio bursts. This scattering process can be effectively modeled as photon diffusion in phase space. In this study, we present a comprehensive comparison between the quasilinear diffusion coefficients and those calculated by ray-tracing the photon trajectories in numerically generated, broadband, isotropic density fluctuation fields in both two-dimensional (2D) and three-dimensional (3D) configurations. 
The comparative analysis demonstrates that for weak scattering, the simulated diffusion coefficients agree \ganjz{well} 
with the quasilinear theoretical predictions. However, when the radio frequency approaches the electron plasma frequency and/or the density fluctuation amplitude becomes significant, photons experience strong scattering. Under such conditions, the quasilinear theory tends to underestimate the scattering strength of photons induced by 2D density fluctuations while overestimating the scattering strength in 3D cases. Furthermore, we implement a group velocity correction to the theoretical diffusion coefficients, based on the effective propagation speed averaged over all test photons. The corrected coefficients provide an accurate quantification of the scattering strength for radio waves propagating through 3D density fluctuations. The physical mechanisms underlying these phenomena are elucidated in the discussion.

\end{abstract}

\keywords{\uat{Solar radio emission}{1522} --- \uat{Radio bursts}{1339} --- \uat{Monte Carlo methods}{2238}} 

\section{Introduction}\label{intro}

Solar storms are violent eruptions of energy in the solar atmosphere. During these storms, the Sun ejects large masses of plasma clouds with strong magnetic field and accelerates charged particles to high energies \citep{aschwanden2006physics}. These energetic particles generate intense electromagnetic radiation, primarily in the form of X-rays, ultraviolet light, and solar radio bursts. Based on their frequency drift rate and morphology in dynamic spectra, solar radio bursts are classified into Type I, II, III, IV, and V bursts \citep{wild1950observatioas,mclean1985solar}, with Type III bursts being the most prolific type of solar radio bursts \citep{reid2014review}. Two coherent emission mechanisms have been proposed for solar radio bursts. The plasma emission mechanism is now the generally accepted model for Type II and Type III burst generation. In this mechanism, high-energy electron beams excite Langmuir waves via bump-on-tail instability, which are then converted into electromagnetic emission near the local plasma frequency or its second harmonics through wave decay or the coalescence of two Langmuir waves \citep{ginzburg1959mechanisms,cairns1995ion,melrose2008quantum,Henri2019JGRsimu,Lee_2019plasmaemission,Ziebell2021Ap&SS,Zhang_2022plasmaemission}. Alternatives, a direct emission mechanism based on the electron cyclotron maser emission (ECME) mechanism has been proposed for Type III bursts \citep{wu2002generation,Yoon2002TypeIII,wu2005altitude,zhao2013solar,wang2015IIIb,chen2017self,chen2021interplanetary}. In the ECME scenario, the electromagnetic wave is initially generated inside a density-depleted flux tube. This wave cannot escape the tube until it reaches an altitude (referred to as the apparent source region),  where the local exterior cutoff frequency (approximately equal to the local plasma frequency in the corona) becomes lower than the wave frequency \citep{wu2002generation}. 

The observed properties of solar radio bursts can be significantly influenced by the propagation effects, \ganjz{as the} solar-terrestrial plasma acts as an inhomogeneous refractive medium for radio waves. 
These effects can be divided into two main types: 1) regular refraction due to large-scale density variations -- the background plasma density decreases gradually with increasing heliocentric distance \citep{newkirk1961solar}, and the resulting density gradient causes radio waves to focus radially during propagation \citep{mann2018tracking}; 2) scattering due to small-scale density fluctuations -- the random changes in the refractive index caused by small-scale plasma turbulence result in random deflection of radio waves. Numerous studies have shown that scattering can lead to: apparent source broadening \citep{sasikumar2017turbulent,murphy2021lofar,Kontar2017NatCo}, changes in radiation directivity \citep{bonnin2008directivity,krupar2018interplanetary}, source position displacements \citep{riddle1974observation,mccauley2018densities,gordovskyy2019frequency,maguire2021lofar,zhang2021parametric,chen2023source}, time delays in radio arrival \citep{steinberg1971coronal,riddle1972effect}, variations in decay time \citep{steinberg1984type,krupar2018interplanetary,chen2020subsecond}, and apparent superluminal movement of radio source \citep{zhang2020interferometric}. Special observational phenomena and fine structures in radio bursts can also be explained through propagation effects. For example, \citet{2009AnGeo..27.3933A} showed that the drifting narrowband fibers observed in Type II bursts are the result of scattering on radio emission; \citet{kuznetsov2020radio} proposed that drift-pair bursts can be formed due to a combination of refraction and anisotropic scattering processes, with the trailing component being the turbulent radio echo.

Ray-tracing simulation is an effective technique for quantitatively investigating the propagation and scattering of radio waves, which was first introduced by \cite{fokker1965coronal} to show that the observed size of type I bursts might be due to scattering. \citet{steinberg1971coronal} and \citet{riddle1972effect} refined this algorithm by incorporating the regular refraction and collision absorption for each individual ray, and applied it to study the effects of scattering on the Type III radio burst characteristics at meter wavelengths.  
 \citet{thejappa2007monte} used a power-law spectrum for the density fluctuations in their simulation program to study the directivity of interplanetary radio bursts, finding that the wide-spread visibility of fundamental and harmonic components can be explained by scattering. Scattering is also responsible for the reduction in brightness temperatures and the increase in the apparent diameters of the quiet Sun at meter-decameter wavelengths \citep{thejappa2008effects}. Observations suggest that density fluctuations in the corona and interplanetary space are usually anisotropic and predominantly perpendicular to the magnetic field lines \citep{Armstrong1990DensObs,kontar2023anisotropic}. Recently, based on the anisotropic scattering model constructed by \citet{kontar2019anisotropic}, simulations have indicated that anisotropic scattering due to statistical inhomogeneity is essential to interpret the properties of observed radio sources \citep{chen2020subsecond,chrysaphi2020first,kuznetsov2020radio, zhang2021parametric, clarkson2025tracing}.   

A key parameter in the ray-tracing simulation of radio wave propagation and scattering is the angular scattering rate, which has been derived using various methods in the literature. The statistical theory for the scattering of radio rays by random fluctuations in plasma density was first introduced by \citet{chandrasekhar1952statistical}. In this framework, the angular scattering rate can be deduced by perturbing the eikonal equation of geometric optics under the approximation of small-angle scattering \citep{hollweg1968statistical, Cairns1998scattring}. Another approach, based on the Hamilton equations for photons, was adopted by \citet{arzner1999radiowave} and \citet{bian2019fokker}, where the scattering of radio waves is described as a diffusion process in wavevector space using a Fokker-Planck equation. \citet{kontar2019anisotropic} further extended the isotropic plasma treatment of \citet{bian2019fokker} to the anisotropic scattering domain. Utilizing quasi-linear theory in plasmas, the wavevector diffusion tensor and the spatial diffusion coefficient for static density fluctuations can be obtained, as briefly outlined in the following Section \ref{theory}. These theoretical results indicate that the scattering coefficients depend on the ratio of the radio wave frequency to the local plasma frequency \(\omega/\omega_{pe}\) and the relative level of density fluctuations \(\epsilon^2={\langle\delta n^2\rangle}/{n^2}\). When the frequency of the radio waves approaches the local plasma frequency, the angular scattering coefficients exhibit a sharp increase. Thus, the fundamental wave may experience strong scattering near its emission source region, while the aforementioned analyses are based on the approximations of small-angle scattering and/or small-amplitude density fluctuations (weak turbulence).

In this paper, we develop an algorithm to trace the trajectories and moving directions of photons in plasmas with statistically homogeneous density fluctuations. Instead of quantitatively investigating the observational properties of the radio emission, we present, for the first time, a comparison between the quasilinear scattering coefficients and those calculated using the ray-tracing code. The paper is structured as follows: the theoretical formulations of angular and spatial diffusion coefficients for radio waves scattered by two-dimensional (2D) density turbulence and by three-dimensional (3D) density fluctuations, respectively, are briefly reviewed in Section \ref{theory}; Section \ref{simulation} details the methodology for constructing our Monte Carlo ray-tracing model; the numerical results of diffusion coefficients in 2D and 3D space are presented in Section \ref{result}, along with discussions and comparisons; and Section \ref{summary} provides the concluding remarks.

\section{Scattering Diffusion Coefficients}
\label{theory}
\subsection{General framework}
Since the inhomogeneity scale of the solar-terrestrial plasma is much larger than the radio wavelength, the effects of diffraction and polarization are negligible. Therefore, the concept of geometric optics is adopted for the propagation of radio waves. Radio waves are treated as photons, whose trajectories are governed by the Hamilton equations \citep[e.g.,][]{arzner1999radiowave}:
\begin{equation}
  \frac{d\mathbf{r}}{dt} = \frac{\partial \omega}{\partial \mathbf{k}}=\mathbf{v}_g,  
  \label{Hamiltonr}
\end{equation}
\begin{equation}
    \frac{d\mathbf{k}}{dt} = -\frac{\partial \omega}{\partial \mathbf{r}}, 
    \label{Hamiltonk}
\end{equation}
where \(\omega\) is the wave frequency, \(\mathbf{k}\) is the wavevector, \(\mathbf{v}_g\) is the wave group velocity, and \(\omega=\omega(\mathbf{r},\mathbf{k})\) is the dispersion relation as a function of position \(\mathbf{r}\) and wavevector \(\mathbf{k}\). For simplicity, the dispersion relation for electromagnetic waves in an unmagnetized plasma is used, 
\begin{equation}
\label{dispersion}
    \omega^2=c^2k^2+\omega_{pe}^2(\mathbf{r}),
\end{equation}
with \(\omega_{pe}=\sqrt{4\pi n(\mathbf{r})e^2/m_e}\) is the plasma frequency, \(n(\mathbf{r})\) is the electron density, and \(e\) and \(m_e\) are the electronic charge and mass, respectively. 

Assuming the plasma to be time independent, i.e. stationary during the transit time of the photons, the electron density \(n(\mathbf{r})\) can be expressed as
\begin{equation}
    n(\mathbf{r})= \bar{n} (\mathbf{r})+\delta n (\mathbf{r}),
\end{equation}
where \(\bar{n}\) represents the locally averaged background density and \(\delta n\) denotes the zero-average density fluctuations. 
In the limit \(\delta n \ll \bar{n}\), the wave Hamiltonian can be decomposed into two parts:
\begin{equation}
    \omega=\bar{\omega}+\delta {\omega} \approx\sqrt{k^2c^2+\bar{\omega}_{pe}^2}+\frac{1}{2}\frac{\bar{\omega}_{pe}^2}{\bar{\omega}}\frac{\delta {n}}{\bar{n}}.
    \label{splitw}
\end{equation}

In the absence of absorption and emission, the number of photons is conserved in the phase space. Using Equations \ref{Hamiltonr}, \ref{Hamiltonk} and \ref{splitw}, the evolution of the phase-space distribution \(N(\mathbf{r},\mathbf{k},t)\) of photons is described by the following wave-kinetic equation \citep{bian2019fokker}
\begin{equation} 
    \frac{\partial N}{\partial t}+\mathbf{v}_g\cdot\frac{\partial N}{\partial \mathbf{r}}+(\bar{f}+\delta{f})\cdot\frac{\partial N}{\partial\mathbf{k}}=0.
    \label{eqwk}
\end{equation}
Here, \(\bar{f}\) is the "refractive force" associated with the relatively smooth change of the density \(\bar n\),
\begin{equation}
    \bar{f}=-\frac{\partial\bar{\omega}}{\partial\mathbf{r}}=-\frac{1}{2\bar{\omega}}\frac{\partial\bar{\omega}_{pe}^2}{\partial\mathbf{r}},
\end{equation}
which leads to regular refraction. On the other hand, \(\delta f\) is a small-scale fluctuating component associated with \(\delta n\),
\begin{equation}
    \delta{f}=-\frac{\partial\delta{\omega}}{\partial\mathbf{r}}=-\frac{1}{2}\frac{\bar{\omega}_{pe}^2}{\bar{\omega}}\frac{\partial}{\partial\mathbf{r}}\left( \frac{\delta{n}(\mathbf{r})}{\bar{n}(\mathbf{r})}\right)
\end{equation}
which results in the aforementioned scattering. 

The scattering term in the wave-kinetic equation \ref{eqwk} induces a diffusion of photons in momentum space. In the quasilinear approximation
, the evolution of the phase-space distribution can be described by a Fokker-Planck equation \citep{arzner1999radiowave,bian2019fokker},
\begin{equation}
     \frac{\partial N}{\partial t}+\mathbf{v}_g\cdot\frac{\partial N}{\partial \mathbf{r}}+\bar{f}\cdot\frac{\partial N}{\partial\mathbf{k}}=\frac{\partial}{\partial k_i}\left[D_{ij}(\mathbf{k})\frac{\partial N}{\partial k_j}\right].
\end{equation}
Here, \(D_{ij}(\mathbf{k})\) is the diffusion tensor in wavevector space, which can be expressed as \citep{bian2019fokker}
\begin{equation}
\label{D_ij}
    D_{ij} = \frac{\pi\bar{\omega}_{pe}^4}{4\omega^2} \int q_i q_j  \frac{\delta n (\mathbf{q})}{\bar{n}} \big| ^2_{\mathbf{q}} \delta(\mathbf{q} \cdot \mathbf{v}_g) d\mathbf{q},
\end{equation}
where \(\mathbf{q}\) is the wavevector of electron density fluctuations, and \(\delta n(\mathbf{q})\) represents the Fourier modes of \(\delta n\). The angular and spatial diffusion coefficients of photons can be derived from Equation \ref{D_ij}.

\subsection{Three-dimensional diffusion coefficients}

For isotropic density fluctuations, due to spherical symmetry, 
the diffusion operator can be simplified as \citep{kontar2019anisotropic}
\begin{equation}
\label{D_mu}
    \frac{\partial}{\partial k_i}\left(D_{ij}\frac{\partial}{\partial k_j}\right)=\frac{\partial}{\partial\mu}\left(\frac{\nu_s}{2}(1-\mu^2)\frac{\partial}{\partial\mu}\right)=\frac{\partial}{\partial\mu}\left(D_{\mu\mu}\frac{\partial}{\partial\mu}\right),
\end{equation}
where \(\mu= \text{cos} \theta\), and \(\theta\) is the polar angle for \(\mathbf{k}\). The scattering frequency is defined as
\begin{equation}
\label{scatteringfrequency}
    \nu_s=\frac{\pi}{8} c\bar{q}\epsilon^2 
          \frac{\bar{\omega}_{pe}^4}{\omega(\omega^2-\bar{\omega}_{pe}^2)^{3/2}},
\end{equation} 
where \(\epsilon^2 =\langle\delta n^2 \rangle / n^2 \), and \(\bar{q}\) 
is the spectrum-averaged mean wavenumber of density fluctuations, and \(c\) is the speed of light in vacuum. 
The angular diffusion coefficient \(D_{\theta\theta}\) is given by:
\begin{equation}
\label{theta_3d}
    D_{\theta\theta} =\frac{d\langle\theta^2\rangle}{2dt}
    =\frac{\pi}{16} c\bar{q}\epsilon^2 
     \frac{\bar{\omega}_{pe}^4}{\omega(\omega^2-\bar{\omega}_{pe}^2)^{3/2}},
\end{equation}
where \({d\langle\theta^2\rangle}/{dt}\) is the angular scattering rate per unit time.
%
Note that there is a constant 1/2 difference between the scattering frequency \( \nu_s \) in Equation \ref{scatteringfrequency} and the definition in \citet{kontar2019anisotropic}. 


As described in the following Section \ref{three-dimensional diffusion}, it is challenging to directly calculate the angular diffusion coefficient from the trajectories of photons in 3D ray-tracing simulation. Instead, we will derive the angular diffusion coefficient from the spatial diffusion coefficient. Angular scattering in the phase space causes photon distributions to approach isotropy. After a sufficiently long propagation time, i.e. \( t \gg \nu_s^{-1} \), the spatial distribution of photons in 3D space can be described by the following diffusion equation \citep{schlegel2020interpolation}:  
\begin{equation}
    \frac{\partial N}{\partial t}=\kappa_{xx}\frac{\partial^2N}{\partial x^2}+\kappa_{yy}\frac{\partial^2N}{\partial y^2}+\kappa_{zz}\frac{\partial^2N}{\partial z^2},
\end{equation}
where the spatial diffusion coefficient, \(\kappa =\kappa_{xx}=\kappa_{yy}=\kappa_{zz}\), is given by
\begin{equation}
\label{spatial_3d}
\begin{split}
    \kappa =\frac{v_g^2}
    {8}\int_{-1}^{+1}d\mu\frac{(1-\mu^2)^2}{D_{\mu\mu}}
    =\frac{v_g^2}{3\nu_s}
    =\frac{8}{3\pi} \frac{c}{\bar{q}\epsilon^2} 
     \frac{(\omega^2-\bar{\omega}_{pe}^2)^{5/2}}{\omega\bar{\omega}_{pe}^4}.
    \end{split}
\end{equation}
Here, \(v_g = c\sqrt{1-\bar{\omega}_{pe}^2/\omega^2}\) is the group velocity of the radio wave. From Equations \ref{theta_3d} and \ref{spatial_3d}, it can be seen that the angular and spatial diffusion coefficients satisfy 
\begin{equation}
\label{anglespace3d}
     D_{\theta\theta} \cdot \kappa = v_g^2/6,
\end{equation}
indicating the mutual derivability between these two coefficients.  

\subsection{Two-dimensional diffusion coefficients}

In this subsection, we examine the propagation of photons in plasmas with a 2D density fluctuation that is isotropic in the perturbation plane. \ganjz{In this case, both the angular and spatial diffusion coefficients can be directly calculated using ray-tracing simulations and compared with the quasilinear theoretical predictions, as detailed in Section \ref{two-dimensional diffusion}.} 

Without loss of generality, the perturbation of the plasma density is assumed to be in the \(x-y\) plane, that is, \(\delta n = \delta n(x,y)\), \ganjz{with photon propagation and scattering restricted to this plane. We introduce a coordinate system with a \(x\)-axis that coincides instantaneously with the direction of the group velocity \(\textbf{v}_g\).} 
From equation \ref{D_ij}, the diffusion tensor \(D_{ij}\) can be rewritten as
\begin{equation}
        D_{ij}(\mathbf{k})=\frac{\pi\bar{\omega}_{pe}^4}{4\omega^2}\int_0^\infty dqq\frac{\delta{n}}{n} \big|^2_{\mathbf{q}}\int_0^{2\pi}d\theta q_iq_j\delta(qv_g\text{cos}\theta),
\end{equation}
where \ganjz{\(\theta\)} denotes the polar angle between the density wave vector \(\mathbf{q}\) and the group velocity \(\mathbf{v}_g\) in the \(x-y\) plane. 

Due to symmetry, the non-diagonal components vanish. The perpendicular and parallel components are given by
\begin{equation}
    D_{\parallel}=D_{xx}=\frac{\pi\bar{\omega}_{pe}^4}{4\omega^2v_g}\int_0^\infty dqq^2\frac{\delta{n}}{n} \big|^2_{\mathbf{q}}\int_0^{2\pi}d\theta \text{cos}^2\theta\delta(\text{cos}\theta)=0,
\end{equation}
\begin{equation}
    \begin{split}
        D_{\perp}=D_{yy} &= \frac{\pi\bar{\omega}_{pe}^4}{4\omega^2v_g} \int_0^\infty dqq^2  \left.\frac{\delta{n}}{n}\right|^2_{\bar{q}}  \int_0^{2\pi} d\theta\sin^2\theta \delta(\cos\theta)  
        \\&=\frac{\bar{\omega}_{pe}^4\bar{q}\epsilon^2}{4\omega c^2k}.
    \end{split}
\end{equation}
Accordingly, the scattering frequency and angular diffusion coefficient for 2D density fluctuations can be directly determined as
\begin{equation}
\label{theta_2d}
    D_{\theta\theta}^{2D}=\nu=\frac{D_{\perp}}{k^2} 
     =\frac{1}{4} c\bar{q}\epsilon^2 \frac{\bar{\omega}_{pe}^4}{\omega(\omega^2-\bar{\omega}_{pe}^2)^{3/2}},
\end{equation}
and the 2D spatial diffusion coefficient is 
\begin{equation}
\label{spatial_2d}
    \kappa_{2D}=\frac{v_g^2}{2\nu}
    =2\frac{c}{\bar{q}\epsilon^2} \frac{(\omega^2-\bar{\omega}_{pe}^2)^{5/2}}{\omega \bar{\omega}_{pe}^4}.
\end{equation}
Similarly to the 3D case, the product of the two diffusion coefficients satisfies
\begin{equation}
\label{anglespace2d}
     D_{\theta\theta}^{2D} \cdot \kappa_{2D} = v_g^2/2,
\end{equation}
as numerically verified in the following Section \ref{two-dimensional diffusion}. In addition,  Equations \ref{theta_3d} and \ref{theta_2d} reveal that, for given values of the radio frequency and the level of density fluctuation, the 2D angular diffusion coefficient is \(4/\pi\) times the 3D angular diffusion coefficient.

\section{Monte Carlo Ray-Tracing Model}
\label{simulation}
The primary objective of this study is to numerically validate the quasilinear diffusion coefficients by tracing and statically analyzing the trajectories of a large number of test photons in a uniform plasma with density fluctuations. The regular refraction of photons due to the large-scale variations in the background plasma density will not be considered in subsequent simulations.

Substituting Equation \ref{dispersion} into Equation \ref{Hamiltonr} and \ref{Hamiltonk}, the test photon trajectories in phase space are calculated by solving the following equations:
\begin{equation}
  \frac{d\mathbf{r}}{dt} = \frac{c^2}{\omega}\mathbf{k}, \ \ \ \ \ \ \ \ \ \ \ \ \
  \label{eq_position}
\end{equation}
\begin{equation}
    \frac{d\mathbf{k}}{dt} = -\frac{\omega_{pe0}^2}{2\omega} \frac{1}{n_0} \frac{\partial \delta n}{\partial \mathbf{r}},
    \label{eq_wavevector}
\end{equation}
where the background plasma density \(n_0\) is a constant, and \(\omega_{pe0} \equiv \bar{\omega}_{pe}=\sqrt{4\pi n_0 e^2/m_e}\). The time-independent density fluctuation is generated by a stochastic Fourier transform:
\begin{equation}
\label{density_model}
    \delta n(\mathbf{r})=\sqrt{2}\epsilon n_0\sum_{j=1}^{N_m}A(q_j)\text{cos}(\mathbf{q}_j\cdot\mathbf{r}+\beta_j),
\end{equation}
where \(N_m\) is the total number of wave modes. \(A(q_j)\), \(\mathbf{q}_j\) and  \(\beta_j\) represent the relative amplitude, the wave vector, and the random phase angle for the \(j\text{th}\) mode, respectively. The propagation directions of the wave modes are randomly distributed in space. With \(\mathbf{q} = q \mathbf{\hat{q}}\), a random wave (unit) vector is chosen as
\begin{equation}
\mathbf{\hat{q}} = 
\begin{pmatrix}
   \hat{q}_x \\ \hat{q}_y \\ \hat{q}_z
\end{pmatrix}
=
\begin{pmatrix}
\sqrt{1-\eta^2}\cos\phi \\
\sqrt{1-\eta^2}\sin\phi \\
\eta
\end{pmatrix}
\end{equation}
with \(\phi \in [0,2\pi]\), \(\eta \in [-1,1]\) for 3D density fluctuations and \(\eta=0\) for 2D density fluctuations.

The sum in Equation \ref{density_model} extends over \(N_m\) logarithmically spaced wavenumbers such that \(\Delta q_j/q_j\) is kept constant. The amplitude function, \(A(q_j)\), is given by
\begin{equation}
    A^2(q_j)=\frac{S(q_j)q_j}{\sum_{n=1}^{N_m}S(q_n)q_n}.
\end{equation}
Here, the density fluctuation spectrum \(S(q)\) is assumed to satisfy an isotropic power-law relation  \citep[e.g.,][]{thejappa2007monte,arzner1999radiowave,kontar2019anisotropic},  
\begin{equation}
S(q) = C_N q^{-(p+d-1)}  \ \ \ \ \ \ \  (q_o < q < q_i),
\end{equation}
 with \(l_o=2\pi/q_o\) and \(l_i=2\pi/q_i\) the outer and inner scales and \(C_N\) the normalization constant. The spectral index \(p\) is chosen as \(5/3\), and \(d\) is the spatial dimension number of density fluctuations, equal to 2 or 3. 

The equations of the photon trajectory are solved using the Runge-Kutta-Fehlberg (RKF45) method \citep{fehlberg1969klassische}. RKF45 automatically adjusts the step size due to the difference between the fourth-order and fifth-order Runge-Kutta solutions at each step, and the maximum tolerated difference is set as \(tol\). The electron density, radio wave frequency, velocity, and length are normalized with respect to the background density \(n_0\), the background plasma frequency \(\omega_{pe0}\), the speed of light \(c\), and the power-weighted mean wavelength of the density fluctuation spectrum \(l_c=2\pi/\bar{q}\), respectively.

\section{Numerical results and discussion}
\label{result}

In the following test photon calculations, the total number of density wave modes, the ratio of the outer scale to the inner scale, and the error control tolerance are set as \(N_m = 300\), \(l_o / l_i = 10\), and \(tol=10^{-5}\), respectively. 
\ganjz{The choice of \(N_m = 300\) represents an optimal compromise between computational accuracy and efficiency, where \(N_m > l_o/l_i\) ensures both statistical reliability in photon behavior and convergence of diffusion coefficients \citep{TD_2013PhPl,schlegel2020interpolation}.}
The tuning parameters for different simulation cases are the density fluctuation level \(\epsilon\) and the frequency ratio \(\omega/\omega_{pe0}\), which determine the values of the scattering coefficients. Each run involves approximately \(10^3\) photons, which are initially randomly distributed in both spatial position and propagation direction. Their initial wavenumbers \(k\) are calculated using the dispersion Equation \ref{dispersion} for a given radio wave frequency \(\omega\). 

\begin{figure}[h]
  \centering
  \begin{subfigure}[b]{0.45\textwidth}
    \includegraphics[width=\linewidth]{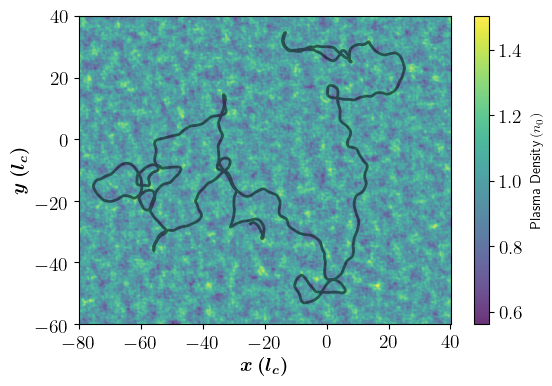}
    \caption{\(\omega/\omega_{pe0}=1.1,\epsilon=0.1\)}
    \label{fig:1a}
  \end{subfigure}
  \hfill 
  \begin{subfigure}[b]{0.45\textwidth}
    \includegraphics[width=\linewidth]{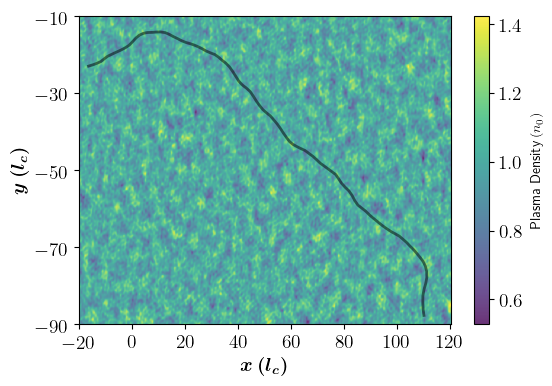}
    \caption{\(\omega/\omega_{pe0}=1.3,\epsilon=0.1\)}
    \label{fig:1b}
  \end{subfigure}
  \caption{Individual simulated photon trajectories (black lines) corresponding to different frequency ratios (\(\omega/\omega_{pe0}\)) and relative level of density fluctuations (\(\epsilon\)). The color shading represents plasma density values.}
  \label{fig:1}
\end{figure}

To illustrate the effects of different scattering strengths on radio wave propagation, Figure \ref{fig:1} displays typical trajectories of two sample photons in plasmas with 2D density fluctuations, calculated from the ray-tracing model. Here, the variations in plasma density and the propagation of the photon are both assumed to be \ganjz{in the} \(x-y\) plane.  
Specifically, Figure \ref{fig:1a} and Figure \ref{fig:1b} present the results under the conditions of strong scattering with \( (\omega/\omega_{pe0}, \epsilon) = (1.1, 0.1) \) and weak scattering with \( (\omega/\omega_{pe0}, \epsilon) = (1.3, 0.1) \), respectively. It is evident that photons tend to travel towards regions of lower electron density, and their propagation directions change randomly due to density fluctuations. In Figure \ref{fig:1a} for strong scattering, photon trajectories exhibit significantly large-angle deflections upon encountering high-density regions, resulting in highly tortuous propagation paths. In contrast, as shown in Figure \ref{fig:1b}, when the wave frequency sufficiently exceeds the plasma frequency, the scattering-induced deflection angles are notably smaller, indicating weak scattering. The photon propagation trajectory remains nearly straight for a longer distance.

\subsection{Two-dimensional diffusion coefficients}
\label{two-dimensional diffusion}

According to \citet{giacalone1999transport}, there are three methods to calculate the diffusion coefficients from the ensemble average of the trajectories of test particle. This work adopts its third method: computing the running diffusion coefficients \citep[e.g.,][]{Qin2002subdiffusive},
\begin{equation}
\label{running_diffusion_coefficient}    \kappa_{\xi\xi}=\lim_{t \to \infty} \frac{\langle \Delta \xi^2 \rangle}{2  t},
\end{equation}
where \(t\) denotes the simulation time, which is required to be much larger than the mean scattering time. Here, \(\Delta \xi\) can be the displacement \(\Delta x\), \(\Delta y\), \(\Delta z\), or the deflection angle \(\Delta \theta\) of the photon propagation direction, and the brackets \(\langle...\rangle\) indicate the ensemble average for all photons. If we get constant values of running diffusion coefficients, we consider to get true diffusion. The simulation time is set to \(200 \nu^{-1}\) to ensure that it is sufficiently long to enter the diffusive regime. 

\begin{figure}[h]
    \centering
    \includegraphics[width=0.8\linewidth]{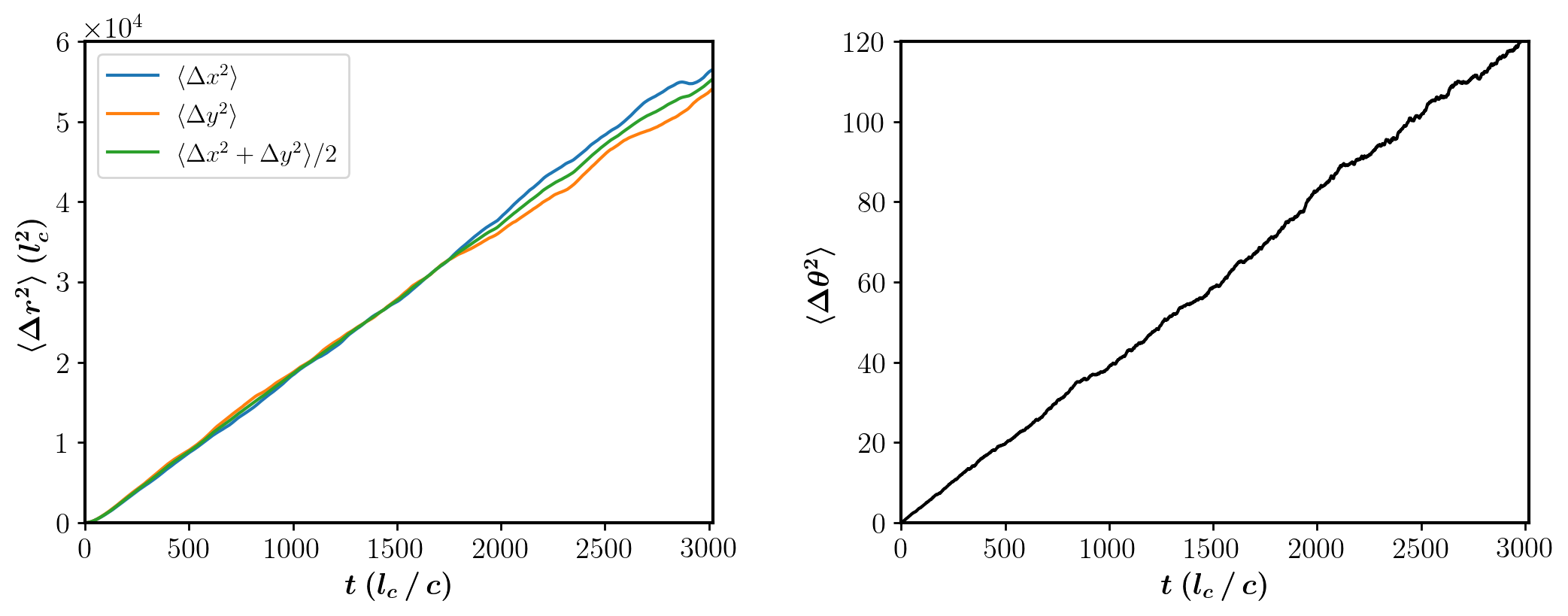}
    \caption{Results of a run involving an ensemble of photons scattered by 2D density fluctuations in the \(x-y\) plane with parameters \((\omega/\omega_{pe0}, \epsilon)=(1.3, 0.1)\). Temporal changes of \(\Delta x^2\), \(\Delta y^2\) and \(\Delta\theta^2\), averaged over all 1000 photons, are plotted.}
    \label{fig:2}
\end{figure}

As an example, Figure \ref{fig:2} shows the time evolution of the mean square displacements and the deflection angle for the case of \((\omega/\omega_{pe0},\epsilon)=(1.3,0.1)\). We can find that \(\langle\Delta x^2\rangle\), \(\langle\Delta y^2\rangle\), and \(\langle\Delta\theta^2\rangle\) all increase almost linearly with time, 
since \(\langle\Delta\xi^2\rangle\propto t\) for a standard diffusion process. In this 2D simulation, because photons are propagating and being deflected only within the \(x-y\) plane, it is convenient to compute the cumulative deflection angle \(\Delta \theta\) of a photon as follows. Let \(\hat{\mathbf{k}}_j\) be the unit vector along the radio wavevector at time step \(t_j\), and \(\hat{\mathbf{n}}\) be a unit vector normal to the \(x-y\) plane, the deflection angle between time steps \(t_j\) and \(t_{j+1}\) is computed as 
\begin{equation}
\delta\theta_j = 
\begin{cases}
+\arccos\left( \hat{\mathbf{k}}_j \cdot \hat{\mathbf{k}}_{j+1} \right), & \text{if } (\hat{\mathbf{k}}_{j} \times \hat{\mathbf{k}}_{j+1}) \cdot \hat{\mathbf{n}} \geq 0. \\
-\arccos\left( \hat{\mathbf{k}}_j \cdot \hat{\mathbf{k}}_{j+1} \right), & \text{if } (\hat{\mathbf{k}}_j \times \hat{\mathbf{k}}_{j+1}) \cdot \hat{\mathbf{n}} < 0.
\end{cases}
\end{equation}
Here, the time-step size, \(\Delta t_j = t_{j+1}-t_j\), is sufficiently small to ensure \(|\delta\theta_j| \ll 1\). The deflection angle \(\delta\theta_j\) is positive for a counterclockwise rotation about \(\hat{\mathbf{n}}\) and negative for a clockwise rotation. So, the cumulative deflection angle up to \(t_n\) is given by 
\begin{equation}
    \Delta\theta(t_n)=\sum_{j=0}^{n-1}\delta\theta_j.
\end{equation}
This approach accurately describes the continuous evolution of the photon's propagation direction and prevents the underestimation of the total deflection angle over long-term propagation.

\begin{figure}[h]
    \centering
    \begin{subfigure}[b]{0.9\textwidth}
        \includegraphics[width=\linewidth]{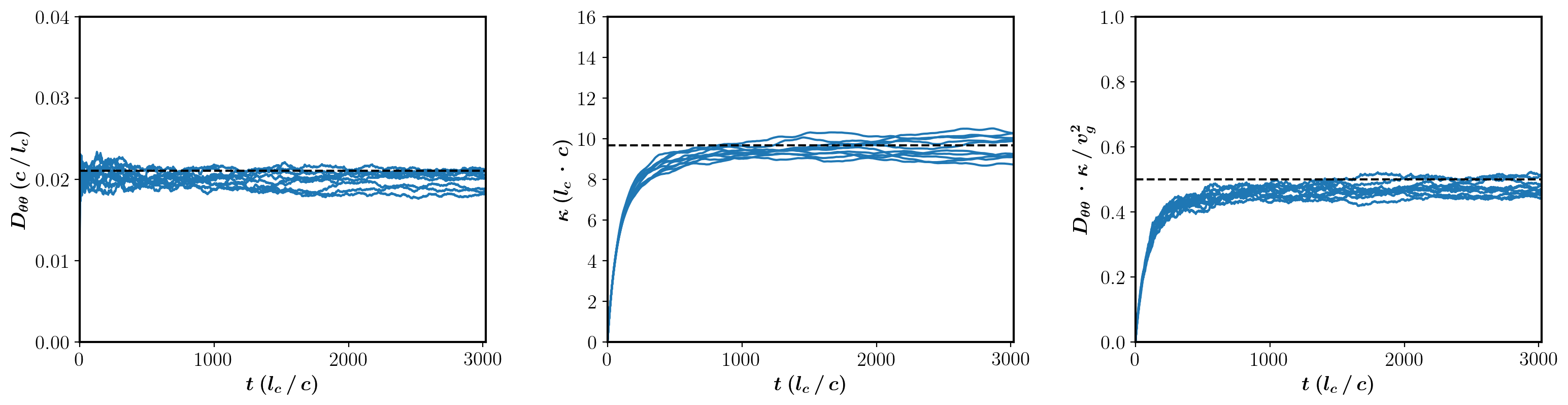}
        \caption{\(\omega/\omega_{pe0}=1.3, \epsilon=0.1\)}
        \label{fig3(a)}
    \end{subfigure}   
    \vspace{0.5cm} 
    \begin{subfigure}[b]{0.9\textwidth}
        \includegraphics[width=\linewidth]{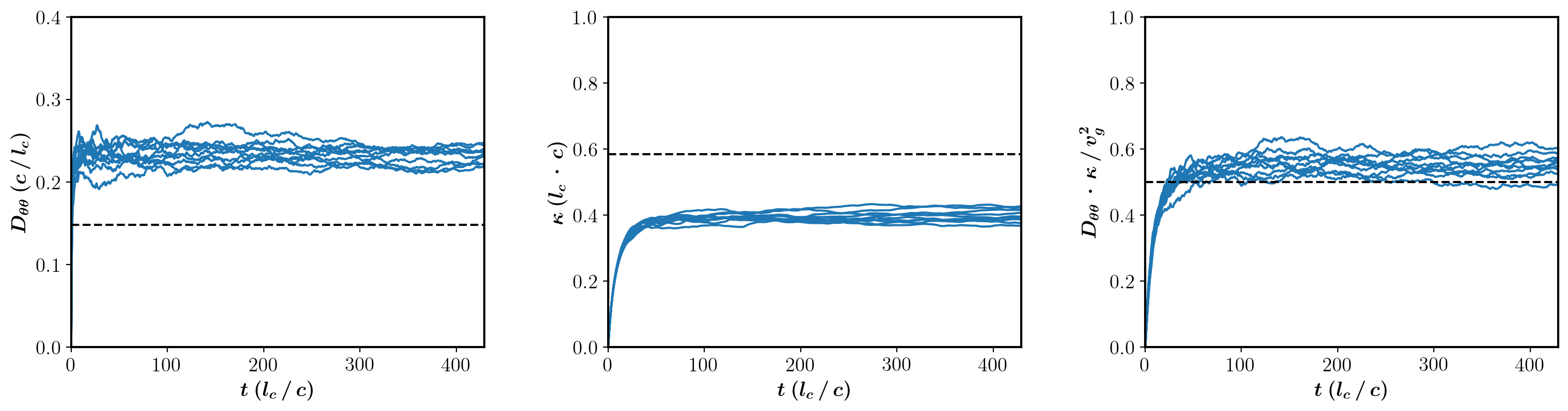}
        \caption{\(\omega/\omega_{pe0}=1.1, \epsilon=0.1\)}
        \label{fig3(b)}
    \end{subfigure} 
    \caption{Diffusion coefficients in 2D density fluctuations obtained from ray-tracing simulations with 1000 photons across 10 different density field realizations for parameter sets \((\omega/\omega_{pe0},\epsilon) = (1.3,0.1) \) and \((1.1, 0.1)\). Here, the angular diffusion coefficient \(D_{\theta\theta}\) (left), the spatial diffusion coefficient \(\kappa\) (middle), \ganjz{and their product} \(D_{\theta\theta}\cdot\kappa/v_{g0}^2\) (right) are plotted. Each blue curve corresponds to the diffusion coefficients calculated from simulation in one realization. The theoretical diffusion coefficients are plotted by black dashed lines. (a) and (b) represent the case of weak scattering and strong scattering, respectively. The last 10\% of the simulation data are used to determine the mean diffusion coefficient.}
    \label{fig3}
\end{figure}

A comparison between simulated and theoretical diffusion coefficients is presented in Figures \ref{fig3(a)} and \ref{fig3(b)} for the parameter sets \((\omega/\omega_{pe0},\epsilon) = (1.3,0.1) \) and \((1.1, 0.1)\), respectively. The simulated diffusion coefficients are calculated according to Equation \ref{running_diffusion_coefficient} for different \(t\).  For each parameter set, 10 independent density field realizations are generated, and the diffusion coefficients are calculated separately for each realization. Given the isotropy of the system, the spatial diffusion coefficient is calculated as \(\kappa = \langle\Delta x^2 + \Delta y^2\rangle/4 t\). Each blue curve corresponds to the diffusion coefficients calculated from the simulation in one realization, and the black dashed line represents the diffusion coefficients derived from quasilinear theory, as given by Equations \ref{theta_2d} and \ref{spatial_2d}. Here, we ensure that \(t \gg \nu^{-1} \), so that the test photon has entered the diffusion regime, where the diffusion coefficients have reached their plateaus and are independent of \(t\). Consequently, only the last 10\% of the mean data is used in the following to determine the simulated diffusion coefficient and to compare it with the theoretical coefficient.

Figure \ref{fig3(a)} represents the results for the condition of weak scattering, where the simulated angular (\(D_{\theta\theta}\)) and spatial (\(\kappa\)) diffusion coefficients are in \ganjz{good} agreement with the theoretical values. 

Figure \ref{fig3(b)} displays the results for the case of strong scattering. Here, the simulated angular diffusion coefficient \(D_{\theta\theta}\) is \textcolor{black}{ \(0.232 \pm 0.003\) (\(c/l_c\)) }, which is about 1.57 times higher than the theoretical value of \textcolor{black}{0.148 (\(c/l_c\))}. The simulated spatial diffusion coefficient \(\kappa\) is \textcolor{black}{\(0.397 \pm 0.006\) (\(cl_c\))}, which is about 1.47 times less than the theoretical value of \textcolor{black}{0.585 (\(cl_c\))}. This indicates that spatial diffusion of radio waves in the 2D plane is significantly suppressed, whereas angular deflections are much stronger than theoretically expected. 

This enhancement of scattering can be interpreted as follows. From Equations \ref{eq_position} and \ref{eq_wavevector}, the propagation equation of radio wave can be rewritten as
\begin{equation}
    \frac{d^2\mathbf{r}}{dt^2}=-\frac{ \omega_{pe0}^2}{2 \omega^2} \frac{c^2}{n_0}\nabla \delta n,
\end{equation}
which means that the electron density fluctuation \(\delta n\) acts as a potential in which the photons move according to classical mechanics \citep{arzner1999radiowave}. The potential is proportional to the relative level of density fluctuation \(\epsilon\) and inversely proportional to the frequency ratio \(\omega /\omega_{pe0}\). One can expect that the photon may be trapped in low-density "well", if the density perturbation is sufficiently high and the radio frequency approaches the plasma frequency. 
In this condition, it is difficult for photons to diffuse outward; instead they undergo frequent large-angle scattering within some confined regions, as shown in Figure \ref{fig:1a}. As a result, the quasilinear approximation breaks down, which necessitates a limit of small-angle scattering or weak turbulence. Consequently, the simulated spatial diffusion coefficient falls below the quasilinear prediction, while the simulated angular diffusion coefficient exceeds the theoretical value.

Furthermore, Equation \ref{anglespace2d} indicates that the product of the angular and spatial diffusion coefficients in 2D density fluctuations, normalized by the square of the group velocity, \((\kappa \cdot D_{\theta\theta}) / v_g^2\), should be equal to 0.5. As shown in the right panels of Figure \ref{fig3}, the simulated value of this ratio consistently converges to 0.5 for both weak and strong scattering, demonstrating its insensitivity to scattering strength.  

\begin{figure} [h]
    \centering
    \includegraphics[width=0.9\linewidth]{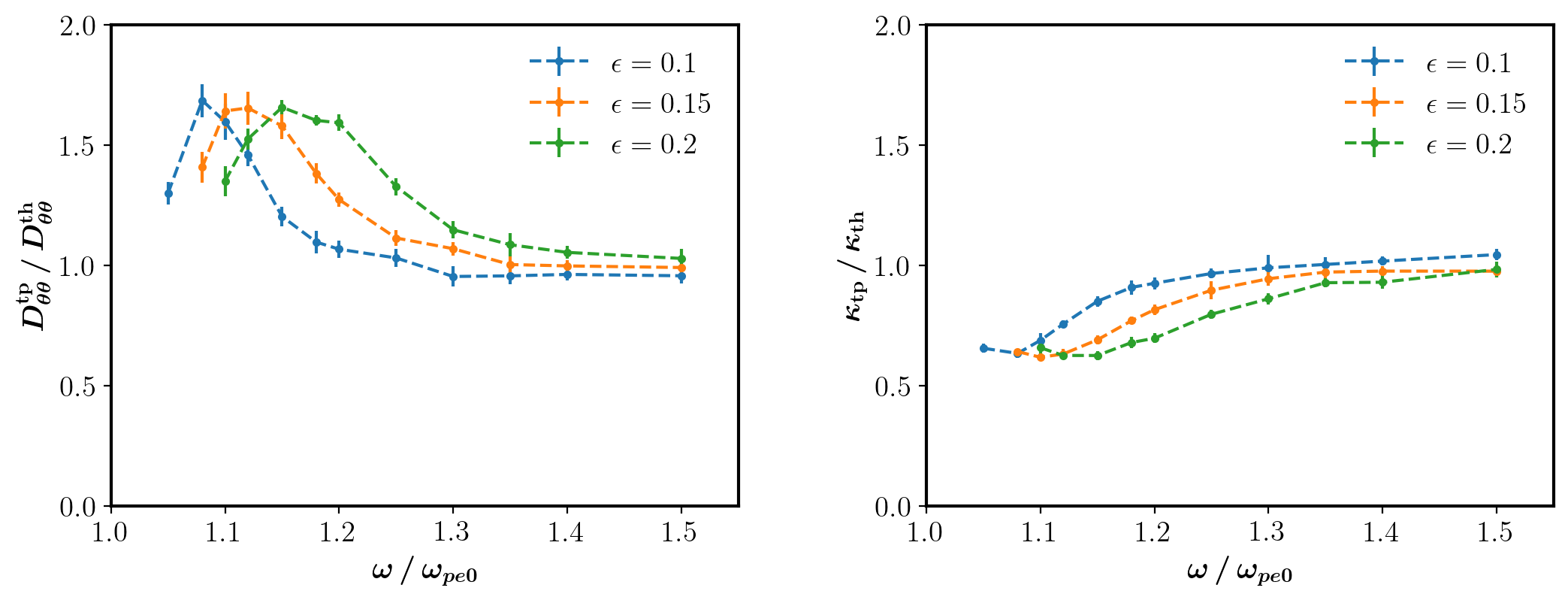}
    \caption{Ratios of simulated-to-theoretical diffusion coefficients as functions of the frequency ratio \(\omega/\omega_{pe0}\) for different density fluctuation amplitudes \(\epsilon\). Left panel: angular diffusion coefficient ratio \(D_{\theta\theta}^{\text{tp}}/D_{\theta\theta}^{\text{th}}\). Right panel: spatial diffusion coefficient ratio \(\kappa_{\text{tp}}/\kappa_{\text{th}}\). Here, \(\epsilon=(0.1, 0.15, 0.2)\) and  \(\omega/\omega_{pe0}\) varying from \(1 + \epsilon/2\) to 1.5. Error bars indicate the standard deviation of 10 independent turbulence realizations.}
    \label{fig4}
\end{figure}

To systematically compare the simulated diffusion coefficients with the predictions of quasilinear theory, we performed ray-tracing simulations over a range of different parameter sets \((\omega/\omega_{pe0},\epsilon)\). Figure \ref{fig4} presents the ratio of the simulated diffusion coefficients (\(D_{\theta\theta}^{\text{tp}}\) or \(\kappa_{\text{tp}}\)) to the theoretical coefficients (\(D_{\theta\theta}^{\text{th}}\) or \(\kappa_{\text{th}}\)). The simulated diffusion coefficient of a given parameter set is calculated as the average value of 10 density fluctuation realizations, and the associated uncertainty is quantified by the standard deviation of these values.

Figure \ref{fig4} shows that when the wave frequency sufficiently exceeds the background plasma frequency, the diffusion coefficient ratio remains approximately equal to one, indicating that quasilinear theory accurately captures the scattering strengths for weak scattering. As the frequency ratio \(\omega/\omega_{pe0}\) decreases or the fluctuation level \(\epsilon\) increases, the ratio of angular diffusion coefficients (\(D_{\theta\theta}^\text{tp} / D_{\theta\theta}^\text{th}\)) first increases and becomes greater than one, while the ratio of spatial diffusion coefficients (\(\kappa_\text{tp} / \kappa_\text{th}\)) falls below one. This behavior can be attributed to the "trapping" effect observed in strong scattering. However, with a further decrease in wave frequency, it is noteworthy that the ratio \(D_{\theta\theta}^\text{tp} / D_{\theta\theta}^\text{th}\) actually decreases. By examining the time dependence of the diffusion coefficients under these conditions, a subdiffusive phenomenon is identified: the diffusion coefficient decreases with the simulation time, rather than remaining constant. A detailed discussion of this subdiffusion is beyond the scope of this paper.

Based on the results shown in Figure \ref{fig4}, an empirical criterion can be proposed: when the radio frequency ratio \(\omega/\omega_{pe0}> 1 + 2 \epsilon\), the simulated diffusion coefficients agree well with the quasilinear diffusion coefficients; however, when \(\omega/\omega_{pe0} < 1 + 2\epsilon\), the weak scattering assumption breaks down and the theoretical coefficients can significantly underestimate the scattering strength of radio waves in 2D density fluctuations. 

\subsection{Three-dimensional diffusion coefficients}
\label{three-dimensional diffusion}

In this section, we calculate the diffusion coefficients of photons in 3D density fluctuations using the ray-tracing code. However, calculating the cumulative deflection angle \(\Delta \theta\) in 3D is challenging. Unlike in 2D simulations, where photons are scattered within the \(x-y\) plane, photons in 3D density fluctuations can be deflected in any direction. This makes it difficult to select a specific plane on which to sum the small incremental deflection angles of the photon's propagation direction, thus complicating the direct determination of the angular diffusion coefficient \(D_{\theta\theta}\) using Equation \ref{running_diffusion_coefficient}. Instead, in 3D simulations, we first determine the spatial diffusion coefficient \(\kappa\) from the photon trajectories and then obtain \(D_{\theta\theta}\) through the relationship \(D_{\theta\theta}=v_g^2/6\kappa\) (Equation \ref{anglespace3d}). The corresponding relation for 2D density fluctuations, that is, Equation \ref{anglespace2d}, has already been confirmed by simulations with various scattering strengths in Section \ref{two-dimensional diffusion}. Taking into account the isotropy of spatial diffusion, \(\kappa\) is calculated by \(\kappa=\langle\Delta x^2+\Delta y^2+\Delta z^2\rangle/6 t\).

\begin{figure}[b]
  \centering

  \begin{subfigure}[b]{0.9\textwidth}
    \centering
    \begin{subfigure}[b]{0.45\textwidth}
      \includegraphics[width=\linewidth]{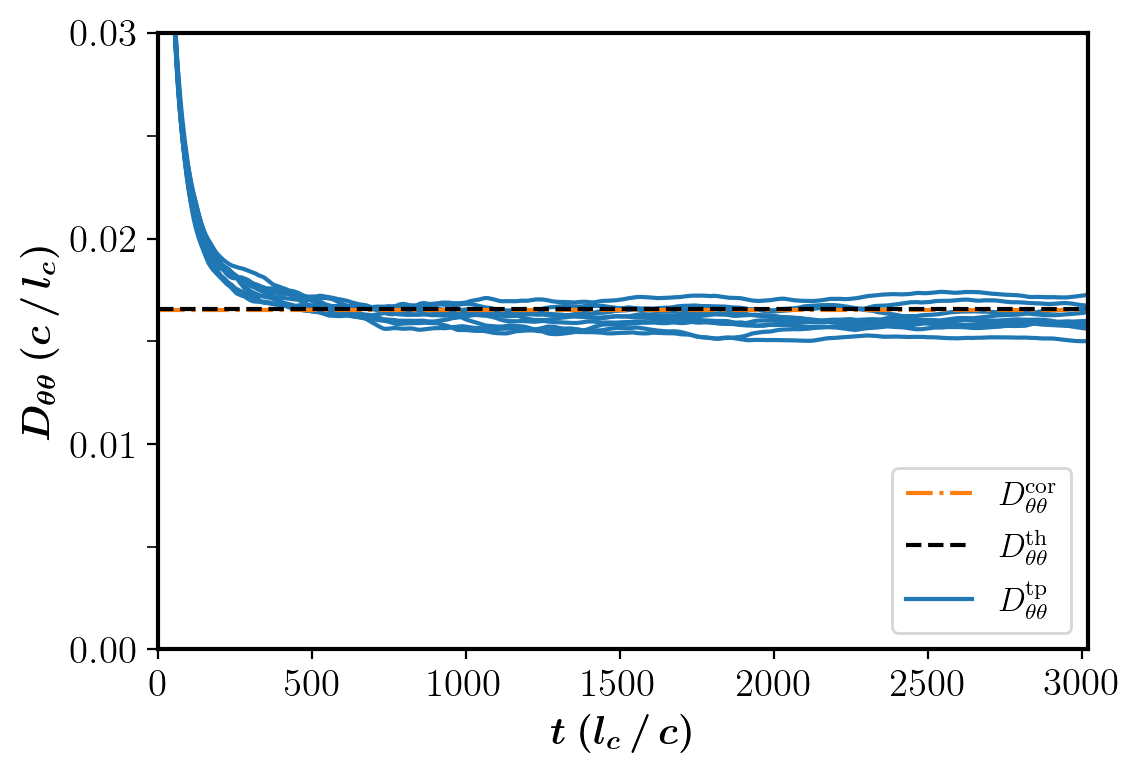}
    \end{subfigure}
    \hfill
    \begin{subfigure}[b]{0.45\textwidth}
      \includegraphics[width=\linewidth]{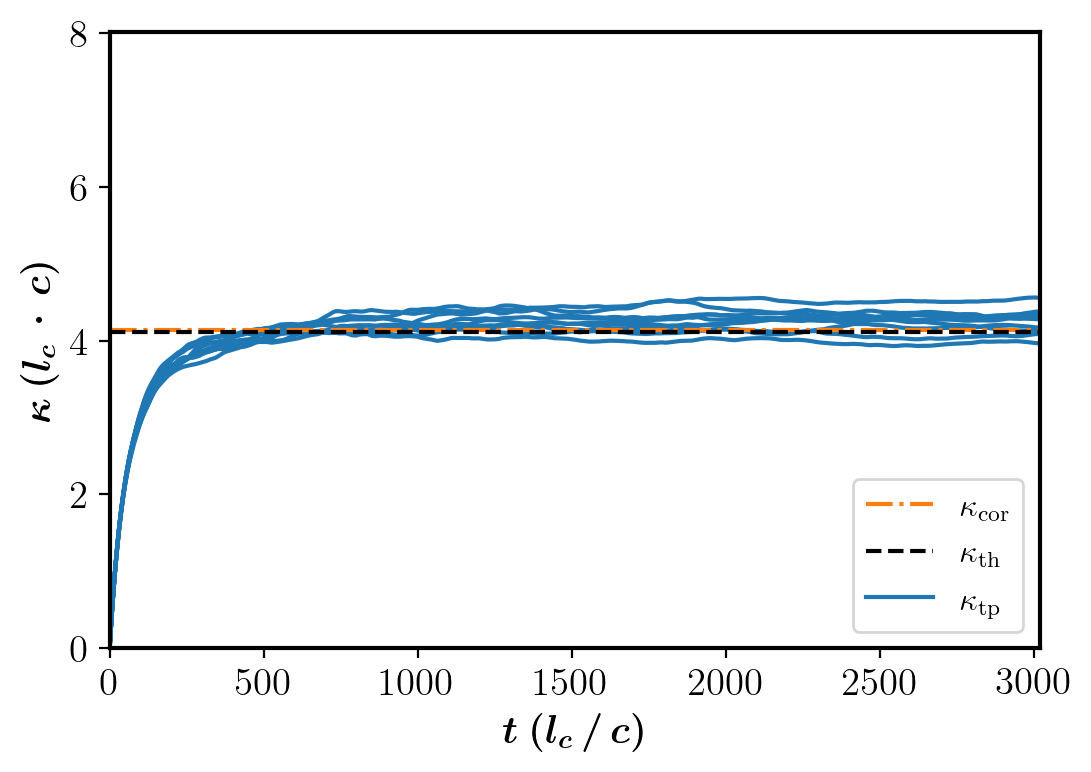}
    \end{subfigure}
    \caption{\(\omega/\omega_{pe0}=1.3,\epsilon=0.1\)}
      \label{fig:5(a)}
  \end{subfigure}

  \vspace{0.5cm}
  
  \begin{subfigure}[b]{0.9\textwidth}
    \centering
    \begin{subfigure}[b]{0.45\textwidth}
      \includegraphics[width=\linewidth]{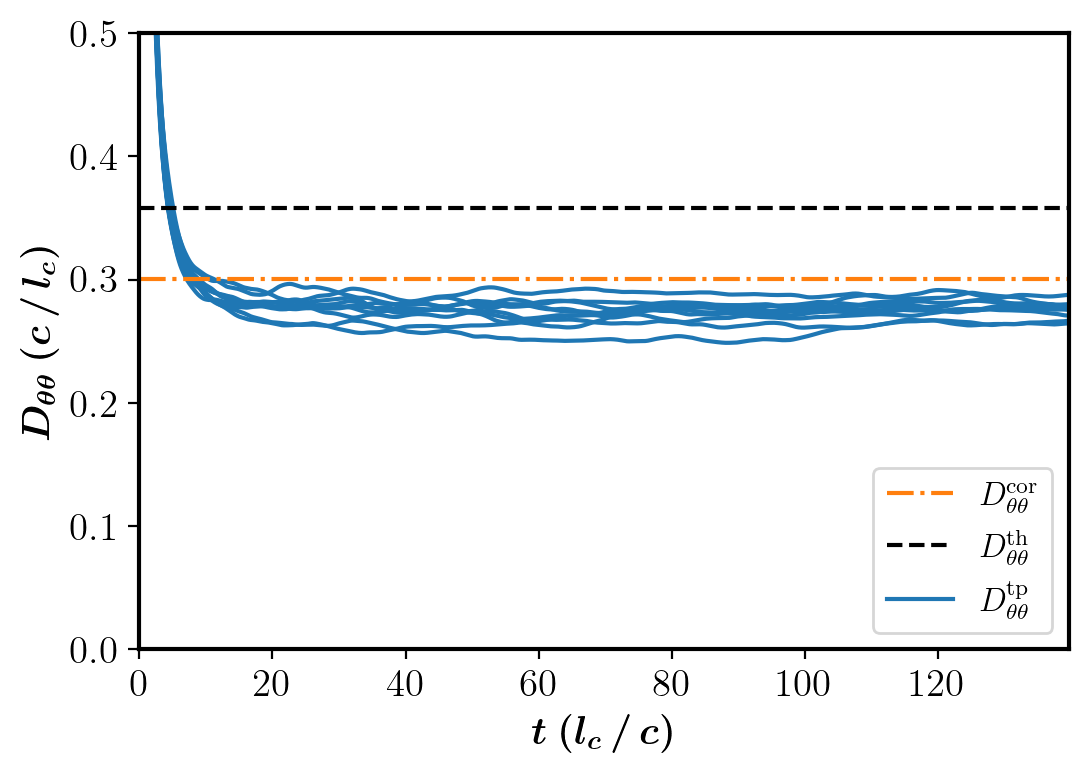}
    \end{subfigure}
    \hfill
    \begin{subfigure}[b]{0.45\textwidth}
      \includegraphics[width=\linewidth]{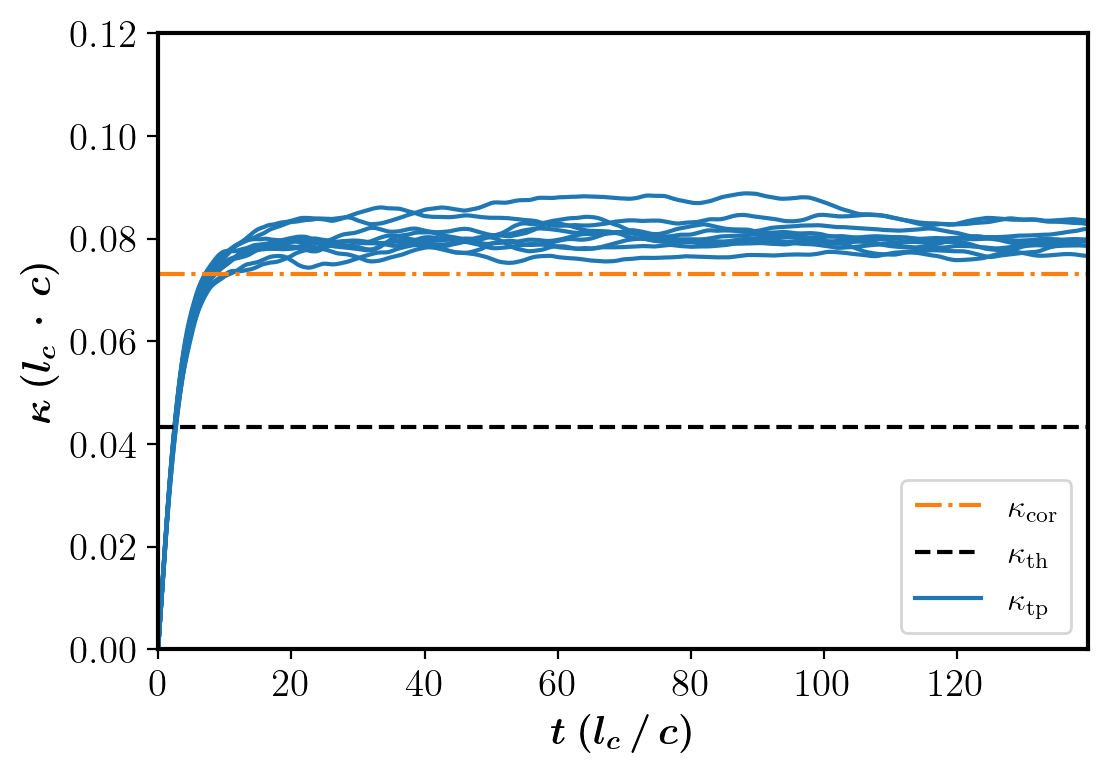}
    \end{subfigure}
    \caption{\(\omega/\omega_{pe0}=1.05,\epsilon=0.1\)}
      \label{fig:5(b)}
  \end{subfigure}

  \caption{Diffusion coefficients in 3D density fluctuations obtained from ray-tracing simulations with 1000 photons across 10 different density field realizations for parameter sets \((\omega/\omega_{pe0},\epsilon)=(1.3,0.1)\) and (\ganjz{1.05},0.1). The angular diffusion coefficient \(D_{\theta\theta}\) (left) and the spatial diffusion coefficient \(\kappa\) (right) are plotted. Each blue curve corresponds to the diffusion coefficients calculated from simulation in one realization, where the last 10\% data are used to determine the mean diffusion coefficient. The theoretical values given by Equations \ref{theta_3d} and \ref{spatial_3d} are plotted with black dashed lines, while the corrected results given by Equations \ref{theta_3d_cor} and \ref{spatial_3d_cor} are plotted with orange dot-dashed lines.}
  \label{fig5}
\end{figure}

Figure \ref{fig5} displays the diffusion coefficients under weak and strong scattering, with parameter pairs \((\omega/\omega_{pe0},\epsilon) = (1.3,0.1)\) and (1.05,0.1) respectively. Similarly to 2D simulations, 10 independent density fields are generated, and the simulated diffusion coefficient calculated from each realization is depicted by a blue curve, while the theoretical predictions (Equations \ref{theta_3d} and \ref{spatial_3d}) are represented by black dashed lines. The orange dot-dashed line indicates the corrected diffusion coefficient, which will be discussed in detail later. It should be noted here that when calculating the simulated \(D_{\theta\theta}\) from the simulated \(\kappa\) using Equation \ref{anglespace3d}, we employed the actual propagation speed averaged across all test photons for each realization, rather than the group velocity \(v_{g0}=c\sqrt{1-\omega_{pe0}^2/\omega^2} \). 

Figure \ref{fig:5(a)} shows \ganjz{good} agreement between the simulated and theoretical diffusion coefficients under weak scattering conditions. However, Figure \ref{fig:5(b)} reveals significant discrepancies for the strong scattering case \((\omega/\omega_{pe0},\epsilon) = (1.05,0.1)\). The simulated spatial diffusion coefficient \(0.0799 \pm 0.0007\) (\(cl_c\)) exceeds the theoretical value of 0.0433 (\(cl_c\)) by a factor of 1.82. Meanwhile, the angular diffusion coefficient \(0.277\pm0.002\) (\(c/l_c\)) is approximately 1.29 times lower than the theoretical prediction of 0.358 (\(c/l_c\)). These results indicate that when the radio frequency approaches the plasma frequency, photon spatial diffusion becomes more efficient than the prediction of quasilinear theory, exhibiting behavior opposite to that observed in the case of 2D strong scattering. 

We propose that this phenomenon arises from a distinctive 3D escape mechanism. Unlike in 2D scenarios where photon deflection is constrained in the \(x-y\) plane, the additional spatial degree of freedom in 3D density fluctuation fields enables photons to circumvent high-density regions more effectively. When encountering dense clumps, photons can adjust their trajectory in the third spatial dimension, preferentially propagating through low-density channels where their velocities typically exceed the group velocity calculated using the background plasma density. This mechanism leads to enhanced spatial diffusion observed in 3D density fluctuations under strong scattering conditions.

\begin{figure}[h]
  \centering
  \begin{subfigure}[b]{0.45\textwidth}
    \includegraphics[width=\linewidth]{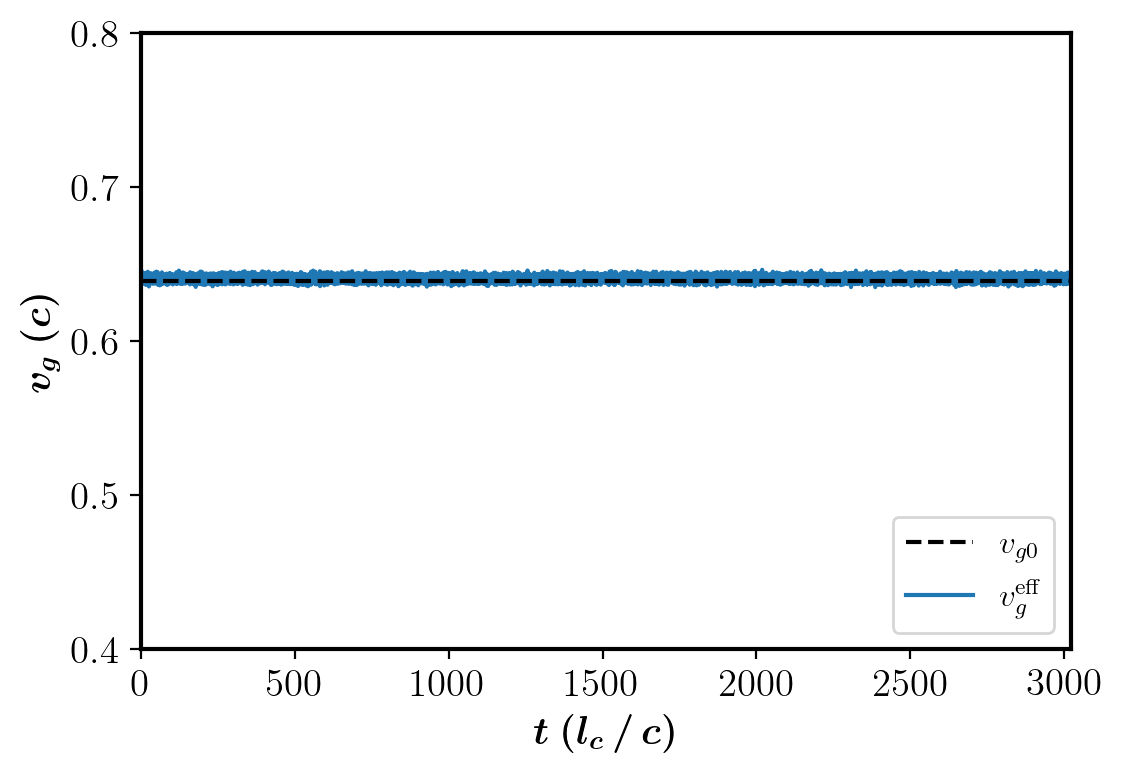}
    \caption{\(\omega/\omega_{pe0}=1.3,\epsilon=0.1\)}
    \label{fig:6a}
  \end{subfigure}
  \hfill 
  \begin{subfigure}[b]{0.45\textwidth}
    \includegraphics[width=\linewidth]{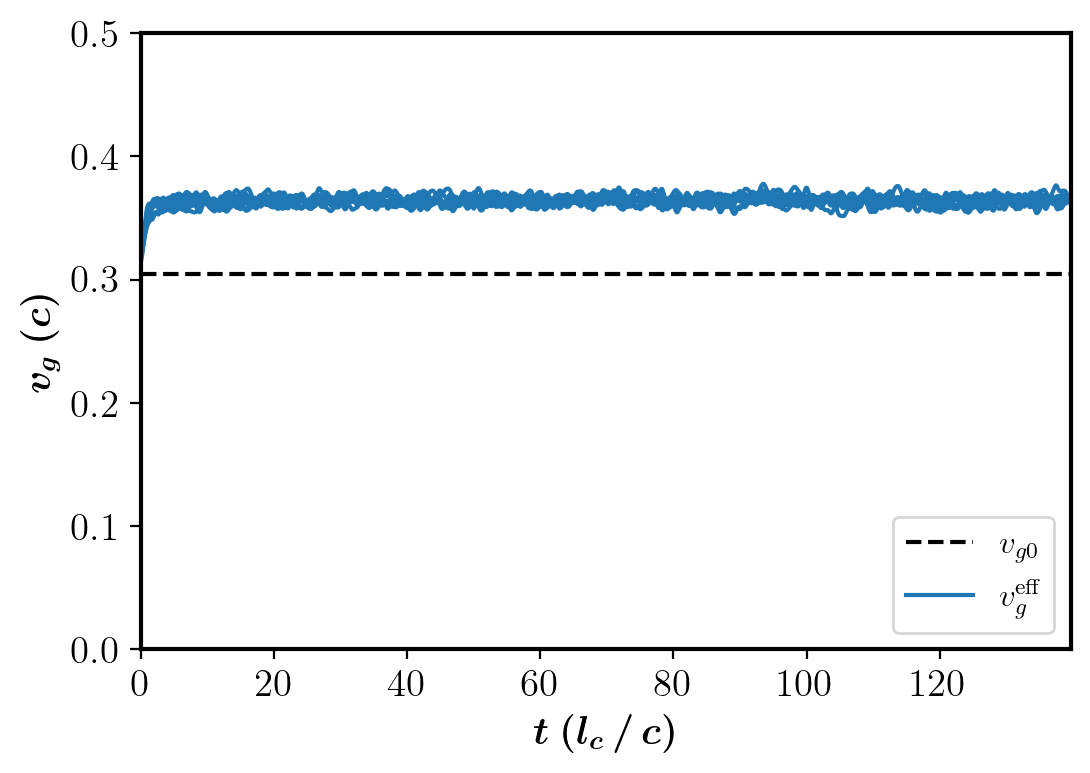}
    \caption{\(\omega/\omega_{pe0}=1.05,\epsilon=0.1\)}
    \label{fig:6b}
  \end{subfigure}
  \caption{Comparison of the real propagation speed \(v_g^\text{eff}\) averaged over all test photons across 10 different realizations of density fluctuation (blue curves) and the group velocity \(v_{g0}\) (black dashed line). (a) and (b) represent the case of weak and strong scattering respectively.}
  \label{fig:6}
\end{figure}

\begin{figure}[t]
    \centering
    \begin{subfigure}[t]{0.9\textwidth}
        \includegraphics[width=\linewidth]{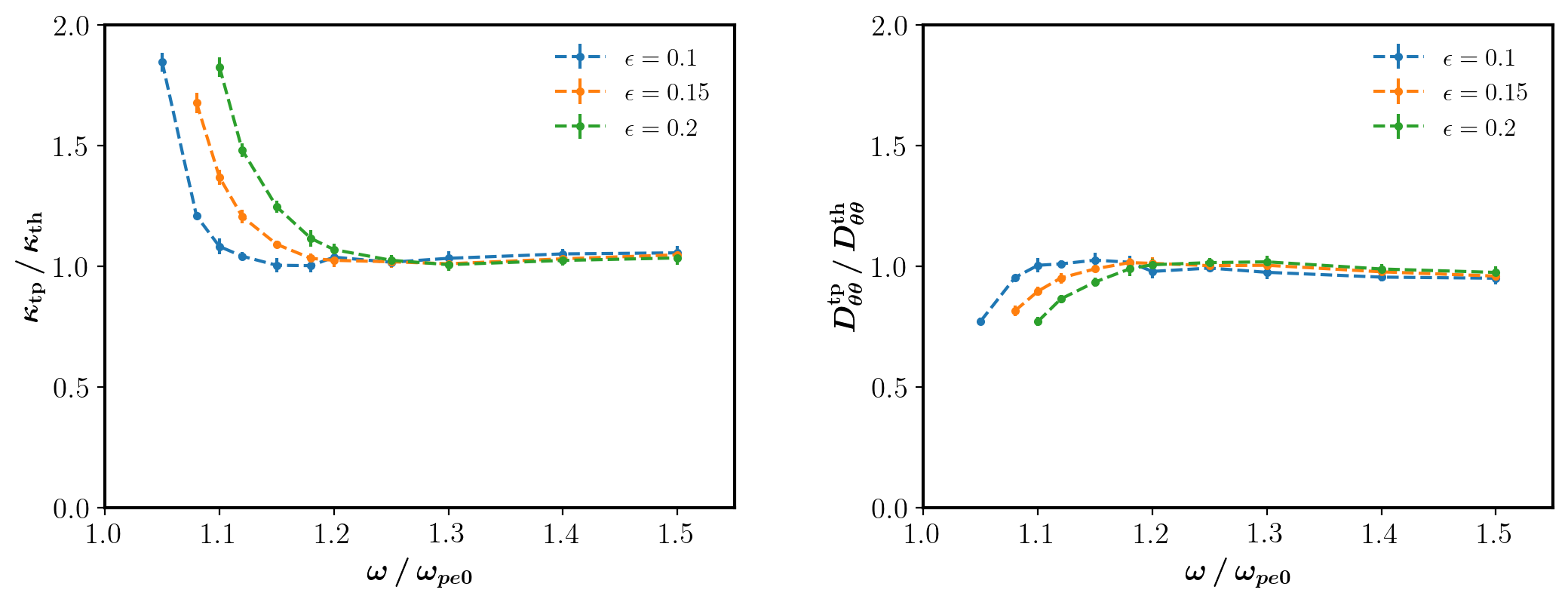}
        \caption{Ratios of simulated-to-theoretical diffusion coefficients \(\kappa_\text{tp}/\kappa_\text{th}\) and \(D_{\theta\theta}^\text{tp} / D_{\theta\theta}^\text{th}\).}
        \label{fig7(a)}
    \end{subfigure}   
    \vspace{0.5cm} 
    \begin{subfigure}[t]{0.9\textwidth}
        \includegraphics[width=\linewidth]{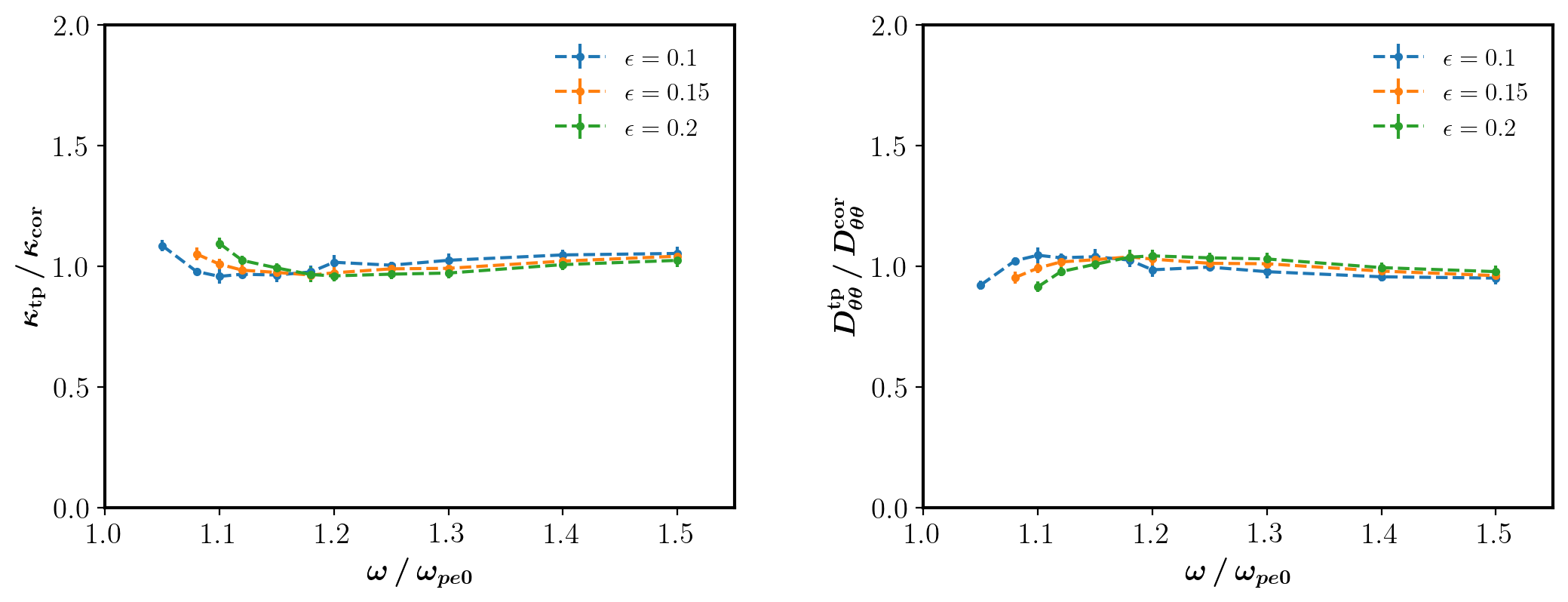}
        \caption{Ratios of simulated-to-corrected diffusion coefficients \(\kappa_\text{tp}/\kappa_\text{cor}\) and \(D_{\theta\theta}^\text{tp} / D_{\theta\theta}^\text{cor}\).}
        \label{fig7(b)}
    \end{subfigure} 
    \caption{Ratios of simulated diffusion coefficients to theoretical coefficients before and after velocity correction in 3D density fluctuations for different values of \((\omega/\omega_{pe0},\epsilon)\). Left panel: spatial diffusion coefficient ratio. Right panel: angular diffusion coefficient ratio. Here,  \(\epsilon = (0.1, 0.15, 0.2)\) and \(\omega/\omega_{pe0}\) varying from 1+\(\epsilon/2\) to 1.5. Error bar indicates the standard deviation across 10 independent turbulent realizations. }
    \label{fig7}
\end{figure}

To validate this interpretation, we compare the effective propagation speed averaged over all test photons \ganjz{\((v_g^{\mathrm{eff}}=\langle k(t)c^2/\omega\rangle)\)} with the background plasma group velocity \(v_{g0}\), as illustrated in Figure \ref{fig:6}. Figure \ref{fig:6a} indicates that for the case of weak scattering,  \(v_g^{\mathrm{eff}}\) remains nearly identical to \(v_{g0}\) throughout the simulation. However, for the case of strong scattering, Figure \ref{fig:6b} shows that \(v_g^{\mathrm{eff}}\) initially equals \(v_{g0}\) but subsequently increases significantly, reaching a time-averaged value of \((0.3639 \pm 0.0001)c\) - approximately 1.19 times higher than the \(v_{g0}\) value of \(0.305 c\). \ganjz{The comparison of \(v_g^{\mathrm{eff}}\) with \(v_{g0}\) shows that, under strong scattering conditions, photons indeed preferentially propagate through low-density channels}.

According to established scattering theory \ganjz{(Equations \ref{D_ij} and \ref{anglespace3d})},  \(D_{\theta\theta}\propto v_g^{-1}\), \(\kappa\propto v_g^3\), we implemented a group velocity correction to the theoretical diffusion coefficients, and the corrected angular and spatial diffusion coefficients are defined as 
\begin{equation}
\label{theta_3d_cor}
    D_{\theta\theta}^{\text{cor}} = D_{\theta\theta}^{\text{th}} \left(\frac{v_g^{\text{eff}}}{v_{g0}} \right)^{-1},
\end{equation}
\begin{equation}
\label{spatial_3d_cor}
    \kappa_{\text{cor}} = \kappa_{\text{th}} \left( \frac{v_g^{\text{eff}}}{v_{g0}} \right)^3,
\end{equation}
where the uncorrected theoretical coefficients of \(D_{\theta\theta}^\text{th}\) and \(\kappa_\text{th}\) are given by Equations \ref{theta_3d}  and \ref{spatial_3d} with \(\bar{\omega}_{pe} = \omega_{pe0}\), respectively. 

The corrected coefficients are represented graphically as orange dot-dashed lines in Figure \ref{fig5}. In the weak scattering regime, where \(v_g^{\mathrm{eff}} \doteq v_{g0}\), the corrected diffusion coefficient lines almost coincide with the theoretical lines. For the strong scattering case of \((\omega/\omega_{pe0},\epsilon)=(1.05,0.1)\), the velocity-corrected diffusion coefficients \(D_{\theta\theta}^\text{cor}\) and \(\kappa_\text{cor}\) are determined to be \(0.301\ (c/l_c\)) and \(0.073\ (c l_c)\), respectively, demonstrating markedly closer alignment with the simulated values compared to the theoretical coefficients, as evidenced in Figure \ref{fig:5(b)}. \ganjz{Although \(D_{\theta\theta}^\text{cor}\) (\(\kappa_\text{cor}\)) remains slightly higher (lower) than the simulated value, 
these findings quantitatively validate} the essential role of velocity correction in strong scattering scenarios.

Furthermore, we performed comprehensive 3D ray-tracing simulations for a series of parameter sets \((\omega/\omega_{pe},\epsilon)\). The comparison between the simulated diffusion coefficients and the theoretical results, both before and after the group velocity correction for various parameter sets, is presented in Figure \ref{fig7}. Analogously to the 2D case, the simulated diffusion coefficient for each parameter pair is calculated as the mean value of 10 independent density fluctuation realizations, with error bars being associated uncertainties, quantified by the corresponding standard deviations.

In Figure \ref{fig7(a)}, when the radio frequency ratio is relatively large, for example, \(\omega/\omega_{pe0} \geq 1.2\), the simulated and uncorrected theoretical diffusion coefficients are in good agreement. 
However, as the radio frequency decreases and/or the density fluctuation level increases, the ratio of the spatial diffusion coefficients gradually ascends, reaching a value about 2, while the corresponding angular diffusion coefficient ratio descends to a value of approximately 0.8. This behavior suggests that, with increasing scattering strength, the quasilinear theory potentially underestimates spatial diffusion and overestimates angular diffusion in 3D density fluctuations, as elucidated previously.

Figure \ref{fig7(b)} illustrates the ratios of the simulated diffusion coefficients to the theoretical coefficients with group velocity correction. It is found that after applying the velocity correction, the diffusion coefficient ratios persist proximately to 1 for all parameter sets. 
These results indicate that the group-velocity-corrected theoretical diffusion coefficients correctly delineate the strength of scattering on radio waves in 3D density fluctuations across the full range of parameters in the simulation.

\section{summary}
\label{summary}
Scattering of radio waves by density fluctuations in solar-terrestrial plasma plays a crucial role in determining the observed properties of solar radio bursts. The scattering (diffusion) coefficients can be derived from the quasilinear theory under the approximation of small-angle scattering or weak turbulence \citep[e.g.,][]{arzner1999radiowave, bian2019fokker}. The strength of scattering is found to depend on the ratio of the wave frequency to the local plasma frequency (\(\omega/\omega_{pe0}\)) and the relative amplitude of density fluctuation (\(\epsilon\)). 
In this investigation, we developed a novel algorithm to trace radio wave propagation trajectories through plasmas with statistically homogeneous density fluctuations, and present for the first time a comprehensive comparison between the quasilinear spatial (angular) diffusion coefficients and those calculated from the ray-tracing code. The main findings are summarized as follows.
\begin{enumerate}
    \item[1)] For relatively high frequency ratios (\(\omega/\omega_{pe0}\)) and low density fluctuation amplitudes where scattering remains weak, the simulated diffusion coefficients are in \ganjz{good} agreement with quasilinear theoretical predictions for both 2D and 3D density fluctuations. 
    In these cases, photon deflection angles remain sufficiently small to validate the approximation of small-angle scattering, confirming the accuracy of quasilinear theory in describing the scattering strength.
    
    \item[2)] When the radio frequency approaches the plasma frequency and/or the density fluctuation amplitude increases significantly, the photons undergo strong scattering. In 2D density fluctuations fields, photons exhibit pronounced trapping behavior in low-density wells. This confinement leads to significantly restricted outward diffusion and frequent large-angle deflection of photons. As a result, the simulated spatial diffusion coefficient falls below the quasilinear theoretical prediction, while the simulated angular diffusion coefficient exceeds the theoretical value. Briefly, the quasilinear theory underestimates the scattering strength under the condition of strong scattering in 2D density fluctuations. 
   
    \item[3)] In 3D density fluctuation fields, the presence of an additional spatial dimension introduces a unique escape mechanism that causes the quasilinear theory to overestimate the scattering strength under strong scattering conditions. The additional degree of freedom enables photons to navigate around high-density regions by adjusting their trajectories along the third spatial dimension. This preferential propagation through low-density channels, where photon velocities are typically larger than the group velocity calculated with the background plasma density, leads to simulated spatial diffusion coefficients that significantly exceed quasilinear predictions. 
    We implemented a group velocity correction to the theoretical diffusion coefficients using the effective propagation speed averaged over all test photons, as defined by Equations \ref{theta_3d_cor} and \ref{spatial_3d_cor}. 
    The modified coefficients correctly determine the strength of scattering on radio waves in 3D density fluctuations across the full range of parameters in the simulation. 
    
\end{enumerate}

The above results reveal that fundamental components of solar radio bursts (with frequencies approaching the local plasma frequency in the source region) can escape from their source region more efficiently than previously anticipated. During outward propagation, these waves preferentially travel in low-density channels, resulting in spatial diffusion rates exceeding theoretical predictions. Therefore, the group velocity-corrected scattering coefficients may enable a more precise simulation of the observational characteristics through ray-tracing techniques. 
For harmonic components (with frequencies approximately twice the local plasma frequency), our simulations confirm that quasilinear theory precisely describes their scattering strength induced by density fluctuations. Finally, this study assumes isotropic density fluctuations; we note that anisotropic scattering may occur in solar-terrestrial regions \citep{robinson1983scattering,kontar2019anisotropic,clarkson2025tracing}.  \ganjz{Typical anisotropic models assume axial symmetry, with dominant fluctuations perpendicular to the ambient magnetic field (anisotropy parameter \(\alpha < 1\)). The 2D density fluctuation model presented in this study represents the extreme anisotropic limit (\(\alpha = 0\)), where density perturbations are confined to the plane perpendicular to the magnetic field. A comprehensive investigation of anisotropic scattering effects with varying \(\alpha\) values warrants future research.} %

\section{Acknowledge}
 The research was supported by the National Nature Science Foundation of China (Grant nos. 42374197 and 41974199) and the Strategic Priority Program of the Chinese Academy of Sciences (Grant No.
XDB41000000). Wang thanks Dr. Gang Qin for useful discussions.

\bibliography{cite}{}
\bibliographystyle{aasjournal}



\end{document}